# Can stress-induced changes in phonon frequencies of ZrSiO$_4$ make it a potential IR spectroscopy-based pressure sensor?

Mubashir Mansoor[1,2], Mehya Mansoor[1,3], Maryam Mansoor[1,4], Zuhal Er[2,5], Kamil Czelej[6*], Mustafa Ürgen[1*]

[1] *Metallurgical and Materials Engineering Department, Istanbul Technical University, Istanbul, Turkey*
[2] *Department of Applied Physics, Istanbul Technical University, Istanbul, Turkey*
[3] *Geological Engineering Department, Istanbul Technical University, Istanbul, Turkey*
[4] *Mining Engineering Department, Istanbul Technical University, Istanbul, Turkey*
[5] *Maritime Faculty, Istanbul Technical University, Istanbul, Turkey*
[6] *Department of complex system modeling, Institute of Theoretical Physics, Faculty of Physics, University of Warsaw, Warszawa, Poland*

A B S T R A C T

Functional materials that can serve as high-pressure transducers are limited, making such sensor material sought after. It has been reported that hydrostatic pressures highly influence Raman shifts of ZrSiO$_4$. Therefore, zirconium silicate has been suggested as a Raman spectroscopic pressure sensor. However, mass applications of a Raman-based sensor technology poses a wide range of challenges. We demonstrate that ZrSiO$_4$ also exhibits pressure-dependent infrared (IR) spectra. Furthermore, the IR peaks of ZrSiO$_4$ are sensitive to shear stresses and non-hydrostatic pressures, making this material a unique sensor for determining a variety of mechanical stresses through IR spectroscopy.

G R A P H I C A L   A B S T R A C T

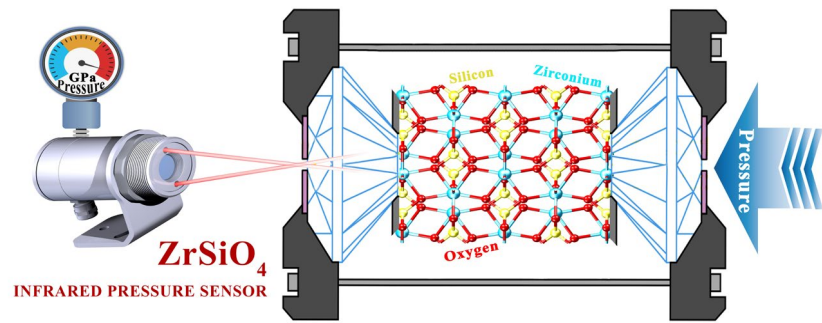

Pressure sensors that can be used for high-pressure processes, as in diamond anvil cells or high pressure high temperature presses (HPHT), are limited [1]. Fluorescence and Raman shifts are among the main material properties used for pressure sensing purposes [2 - 4]. Schmidt et al. [5] have demonstrated the capacity of ZrSiO$_4$ (zircon) as an efficient pressure and temperature sensor under hydrostatic conditions. The pressure dependence of Raman shifts in zircon has even made it possible to use ZrSiO$_4$ inclusions in natural minerals for thermobarometric analysis [6, 7]. The tetragonal structure of zircon may make it a suitable pressure sensor for determining the different types of stresses acting on it, if the variations of Raman shifts arising from strain-dependent phonon frequencies (due to lattice parameter shifts [8]) are large enough. Thus, making it possible to use this material for sensing shear stresses and uniaxial stress components through spectroscopic means.

Table 1. Lattice parameters (a,b,c), density (ρ), change in formation energy (ΔE$^f$), and band gap (ΔE$_g$) at zero Kelvin (referenced to P = 0 GPa), as a function of hydrostatic pressures.

| Pressure (GPa) | a,b (Å) | c (Å) | ρ (g/cm$^3$) | ΔE$^f$ (eV) | ΔE$_g$ (eV) |
|---|---|---|---|---|---|
| 0.0 | 6.673 | 6.028 | 4.536 | 0 | 0 |
| 1.0 | 6.660 | 6.022 | 4.559 | + 0.001 | + 0.028 |
| 2.0 | 6.647 | 6.016 | 4.581 | + 0.005 | + 0.056 |
| 3.0 | 6.635 | 6.010 | 4.602 | + 0.011 | + 0.083 |
| 4.0 | 6.623 | 6.004 | 4.623 | + 0.019 | + 0.109 |
| 5.0 | 6.612 | 5.998 | 4.643 | + 0.028 | + 0.134 |
| 7.5 | 6.585 | 5.983 | 4.693 | + 0.054 | + 0.197 |
| 10.0 | 6.560 | 5.968 | 4.741 | + 0.091 | + 0.256 |
| 15.0 | 6.513 | 5.939 | 4.832 | + 0.190 | + 0.366 |

$ZrSiO_4$ is a refractory ceramic, stable up to approximately 1700 °C [9] and exhibits pressure stability up to 8 GPa (at 298 K) that increases with temperature [10]. Application of higher pressures will initiate a phase transformation to reidite (the high-pressure polymorph of $ZrSiO_4$) [10]. However, studies in the literature show the usability of this material during short-term exposures under pressures well beyond the stable pressure range [11].

Although the sensitivity of Raman spectroscopy can be in principle higher than infrared, the higher cost of in-situ Raman spectroscopy and wide availability of IR-based technologies has prompted us to ask whether IR spectra of zircon can be used for industrial pressure and stress sensing applications. Therefore, we aimed to analyze the pressure dependence of zircon's phonon frequencies through ab-initio calculations and plot the relevant IR and Raman spectra to evaluate the applicability of IR spectroscopy on $ZrSiO_4$ for pressure sensing applications.

We applied spin-polarized density functional theory (SP-DFT) to evaluate phonon frequencies' pressure dependence. Generalized gradient approximation was applied for the exchange-correlation functional, as parametrized by Perdew-Burke-Ernzernhof (GGA-PBE) [13]. The primitive zircon cell was fully relaxed for each hydrostatic pressure until Hellmann-Feynman forces dropped below 1 meV/Å, with a self-consistency energy convergence criterion of $10^{-7}$ eV. The calculations were carried out under periodic boundary conditions, using the supercell approach, with a plane wave energy cutoff of 500 eV, k-point spacing of 0.4 Å$^{-1}$, and Gaussian smearing of 0.05 eV. The projector augmented wave (PAW) method was applied [14] as implemented in VASP code [15]. Cell parameters were varied in the case of non-hydrostatic pressures, and the corresponding stress tensor components were calculated. The phonon dispersion, IR spectra, and Raman shifts were determined using MedeA-Phonon [16] by applying the computational approach proposed by Parlinski, Li, and Kawazoe [17]. The relevant scaling factors for IR spectra and Raman shifts were evaluated based on the unstressed structures.

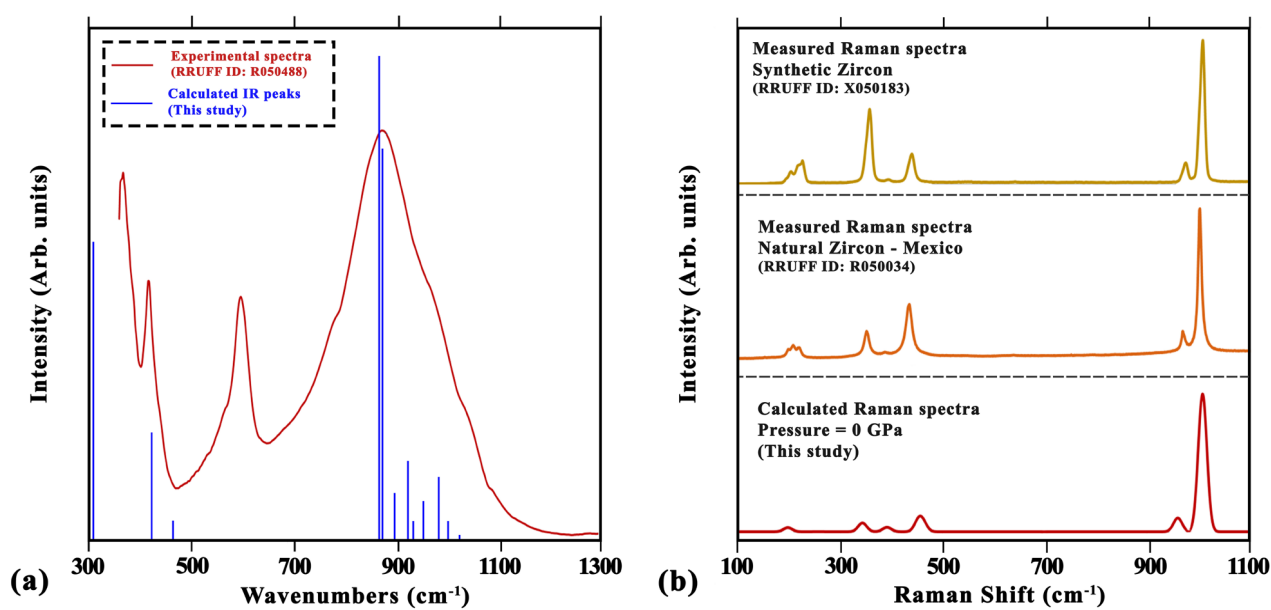

Fig. 1. The IR spectra (a) and Raman shifts (b) of zircon are retrieved from the RRUFF database [19] and compared to our computational output for P = 0 GPa. The methodology employed in this study can successfully predict the IR and Raman spectra.

The calculated lattice parameters, density, changes in formation energy and band gap as a function of hydrostatic pressure are listed in table 1. The lattice parameters agree with the experimental findings of Ehlers et al. [18], with variations of less than 1 %. The computational IR and Raman spectra are benchmarked with the available experimental data from the RRUFF database [19] (Fig. 1). As can be seen, the relative positions of the Raman shifts and IR peaks are well reproduced theoretically, and the relative intensities of the primary phonon mode in both Raman and IR spectra are correctly assigned. Our calculated IR spectrum does not exhibit absorption near 600 cm$^{-1}$ as presented in the experimental data. Therefore, we hypothesize that this IR peak is most probably defect-related.

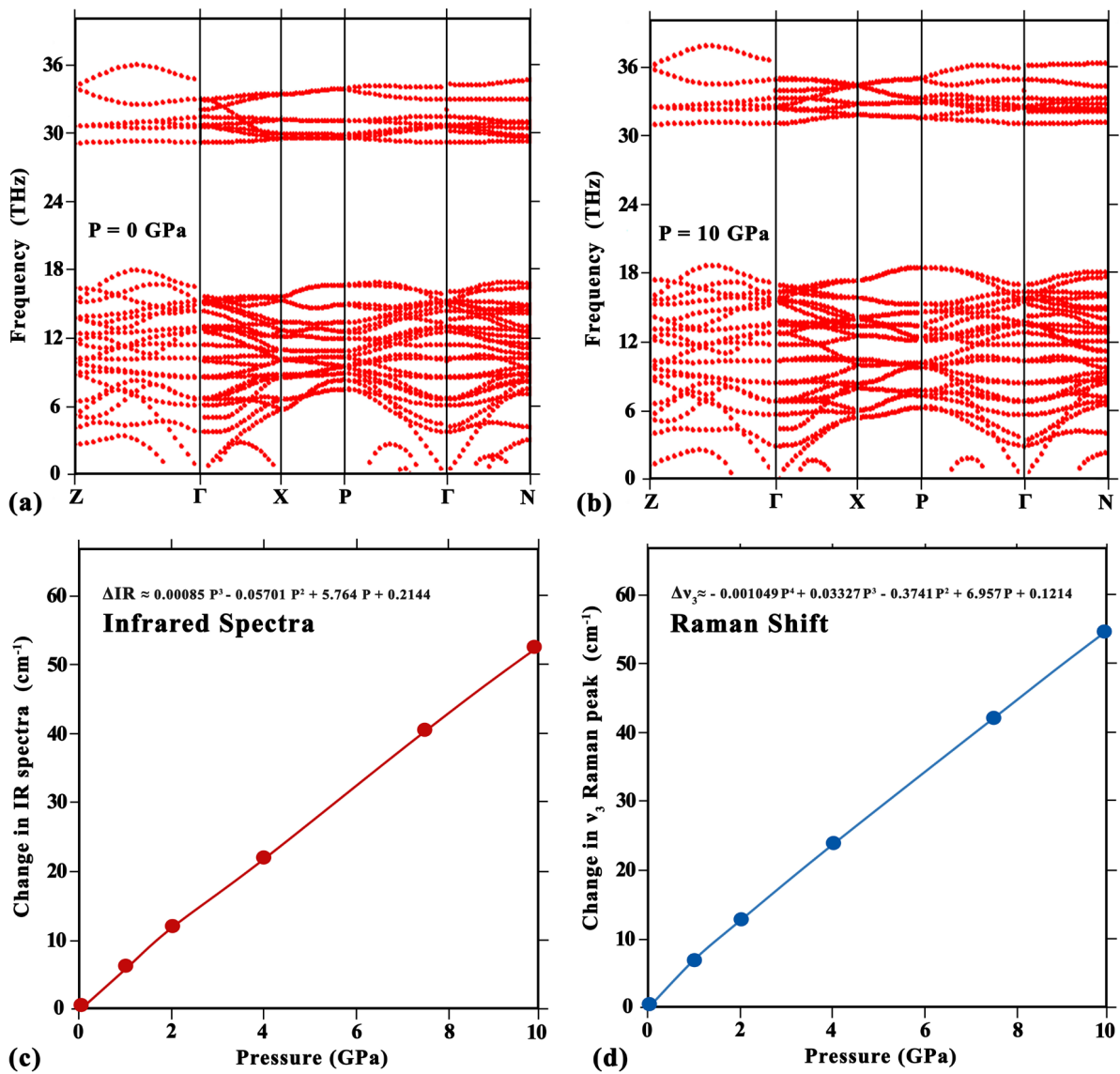

Fig. 2. The phonon dispersions under hydrostatic pressures of P = 0 GPa (a) and P = 10 GPa (b) show a nearly linear shift towards higher frequencies with pressure. The 870 cm$^{-1}$ IR peak (at P = 0 GPa) exhibits pressure dependence (c), as in the $v_3$ Raman peak of zircon (1008.5 cm$^{-1}$ at P = 0 GPa) (d). An increase in pressure increases the Raman shift while decreasing the intensity of the $v_3$ peak. The variations in low-pressure ranges are not linear; therefore, we have used polynomials of higher degrees for curve fitting. The fitting equations are valid for up to 15 GPa.

We find that IR spectra of zircon also show pressure-dependent changes, similar in magnitude with the changes in Raman shifts, demonstrating the possibility to use zircon as an infrared pressure sensor. The changes in phonon dispersions are presented in Fig 2. Almost all phonon frequencies are shifted up with hydrostatic pressure in an approximately linear manner; thus, causing significant changes in the calculated Raman and IR spectra of zircon. The calculated changes in Raman shifts are in close agreement with the experimentally measured [5] and computationally predicted values [12]. The pressure dependence of 870 cm$^{-1}$ IR peak and $\nu_3$ Raman shift (1008.5 cm$^{-1}$ at 0 GPa) under hydrostatic conditions are shown in Figures 2c and 2d.

Additionally, our results indicate that the changes in Raman shifts and IR spectra are significantly different under non-hydrostatic stresses or pressures (Fig. 3). Thus, simultaneous tracking of various vibrational modes of ZrSiO$_4$ can be a clever strategy for measuring uniaxial stresses through IR and Raman spectroscopy, as the spectroscopic shift for each vibrational mode is uniquely correlated to the type of stress tensor applied.

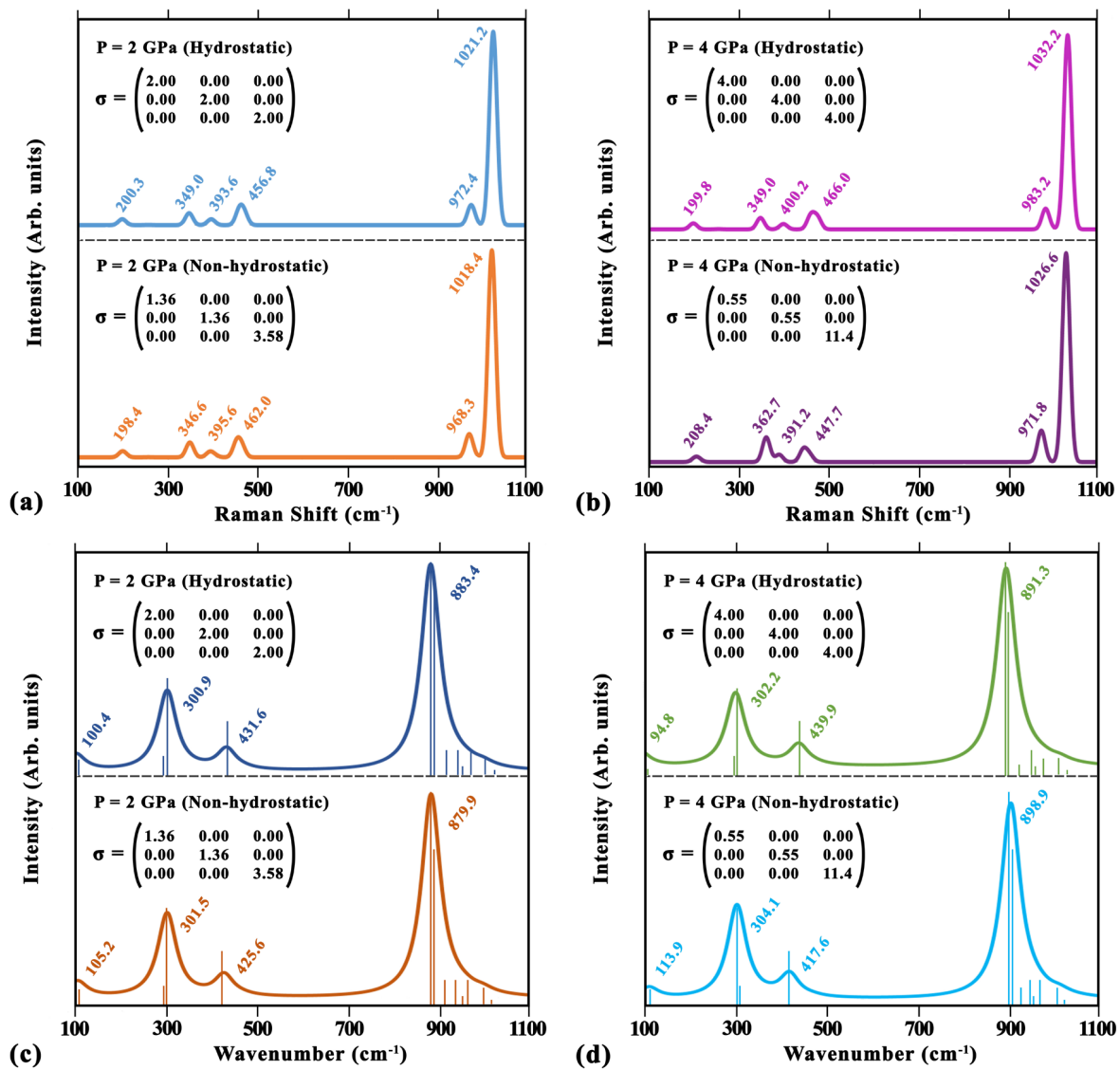

Fig. 3. A comparison of the calculated Raman shifts (a, b) and IR spectra (c, d) under hydrostatic and non-hydrostatic pressures of P = 2 GPa (a, c), and P = 4 GPa (b, d) shows the importance of the applied stress tensors. The positive values in σ tensors are compressive stresses in GPa, with the diagonal values being σ$_{xx}$, σ$_{yy}$, and σ$_{zz}$ respectively. The Raman and IR peaks under non-hydrostatic stresses deviate significantly from the expected hydrostatic calibrations. A Gaussian (Lorentzian) fitting is applied for the Raman (IR) spectra.

In conclusion, we have demonstrated that zircon exhibits significant changes in its IR spectrum as a function of pressure, a behavior that is somewhat similar to the changes in Raman shifts with respect to pressure. Thus, making zircon a functional material for pressure sensing purposes through infrared spectroscopy. Moreover, the Raman shifts and IR spectra are highly sensitive to the stress tensor applied. Therefore, diligent calibration is required in the case of non-hydrostatic pressures when using zircon as a sensor material.

**Acknowledgments**

We are thankful for the fruitful communications with Dr. Rachelle Turnier, Professor John Valley, Professor Özgül Keleş, and Dr. René Windiks. The computational supports of the National Center for High Performance Computing of the Republic of Turkey (UHeM) under Grant No. 1008852020 and the High Performance Computing Facilities of the Interdisciplinary Centre for Mathematical, and Computational Modeling (ICM) of the University of Warsaw under Grant No. GB79-16 are highly appreciated.

**References**

[1] F. Datchi, A. Dewaele, P. Loubeyre, R. Letoullec, Y. Le Godec, B. Canny, Optical pressure sensors for high-pressure–high-temperature studies in a Diamond Anvil Cell, High Pressure Research. 27 (2007) 447–463. doi:10.1080/08957950701659593.

[2] V.G. Baonza, M. Taravillo, A. Arencibia, M. Cáceres, J. Núñez, Diamond as pressure sensor in high-pressure Raman spectroscopy using Sapphire and other GEM Anvil Cells, Journal of Raman Spectroscopy. 34 (2003) 264–270. doi:10.1002/jrs.998.

[3] J.D. Barnett, S. Block, G.J. Piermarini, An optical fluorescence system for quantitative pressure measurement in the diamond-anvil cell, Review of Scientific Instruments. 44 (1973) 1–9. doi:10.1063/1.1685943.

[4] S.G. Lyapin, I.D. Ilichev, A.P. Novikov, V.A. Davydov, V.N. Agafonov, Study of optical properties of the NV and SIV centres in diamond at high pressures, Nanosystems: Physics, Chemistry, Mathematics. (2018) 55–57. doi:10.17586/2220-8054-2018-9-1-55-57.

[5] C. Schmidt, M. Steele-MacInnis, A. Watenphul, M. Wilke, Calibration of zircon as a Raman spectroscopic pressure sensor to high temperatures and application to water-silicate melt systems, American Mineralogist. 98 (2013) 643–650. doi:10.2138/am.2013.4143.

[6] M.L. Mazzucchelli, R.J. Angel, M. Alvaro, EntraPT: An online platform for elastic geothermobarometry, American Mineralogist. 106 (2021) 830–837. doi:10.2138/am-2021-7693ccbyncnd.

[7] R.B. Turnier, J.W. Valley, A. Palke, Raman spectra of zircon inclusions in Sapphire, Geological Society of America Abstracts with Programs. (2018). doi:10.1130/abs/2018am-322879.

[8] C. Stangarone, R.J. Angel, M. Prencipe, N. Campomenosi, B. Mihailova, M. Alvaro, Measurement of strains in zircon inclusions by Raman spectroscopy, European Journal of Mineralogy. 31 (2019) 685–694. doi:10.1127/ejm/2019/0031-2851.

[9] L. Liu, Z. Ma, Z. Yan, S. Zhu, L. Gao, The ZrO2 formation in ZrB2/SiC composite irradiated by Laser, Materials. 8 (2015) 8745–8750. doi:10.3390/ma8125475.

[10] M. Akaogi, S. Hashimoto, H. Kojitani, Thermodynamic properties of ZrSiO4 zircon and reidite and of cotunnite-type ZrO2 with application to high-pressure high-temperature phase relations in ZrSiO4, Physics of the Earth and Planetary Interiors. 281 (2018) 1–7. doi:10.1016/j.pepi.2018.05.001.

[11] A. Gucsik, M. Zhang, C. Koeberl, E.K. Salje, S.A. Redfern, J.M. Pruneda, Infrared & Raman spectra of ZrSiO4 experimentally shocked at high pressures, Mineralogical Magazine. 68 (2004) 801–811. doi:10.1180/0026461046850220.

[12] N. Sheremetyeva, D.J. Cherniak, E.B. Watson, V. Meunier, Effect of pressure on the Raman-active modes of zircon (ZrSiO4): A first-principles study, Physics and Chemistry of Minerals. 45 (2017) 173–184. doi:10.1007/s00269-017-0906-1.

[13] J.P. Perdew, K. Burke, M. Ernzerhof, Generalized gradient approximation made simple, Physical Review Letters. 77 (1996) 3865–3868. doi:10.1103/physrevlett.77.3865.

[14] P.E. Blöchl, Projector augmented-wave method, Physical Review B. 50 (1994) 17953–17979. doi:10.1103/physrevb.50.17953.

[15] G. Kresse, D. Joubert, From Ultrasoft pseudopotentials to the projector augmented-wave method, Physical Review B. 59 (1999) 1758–1775. doi:10.1103/physrevb.59.1758.

[16] MedeA version 3.2.2; MedeA is a registered trademark of Materials Design, Inc., San Diego, USA.

[17] K. Parlinski, Z.Q. Li, Y. Kawazoe, First-principles determination of the soft mode in CubicZrO2, Physical Review Letters. 78 (1997) 4063–4066. doi:10.1103/physrevlett.78.4063.

[18] A.M. Ehlers, G. Zaffiro, R.J. Angel, T. Boffa-Ballaran, M.A. Carpenter, M. Alvaro, et al., Thermoelastic properties of zircon: Implications for geothermobarometry, American Mineralogist. 107 (2022) 74–81. doi:10.2138/am-2021-7731.

[19] B. Lafuente, R. T. Downs, H. Yang, N. Stone, The power of databases: the RRUFF project. In: Highlights in Mineralogical Crystallography, T Armbruster and R M Danisi, eds. Berlin, Germany, W. De Gruyter, (2015) pp 1-30